\documentclass[journal]{IEEEtran}

\usepackage{graphicx}
\usepackage{color}
\usepackage{placeins}
\usepackage{float}
\usepackage{tabularx,colortbl}
\usepackage{amssymb}
\usepackage{amsthm}
\usepackage{cite}
\usepackage{amsmath}
\usepackage{makecell}

\usepackage{caption2}

\usepackage{algorithm}
\usepackage{algpseudocode}

\theoremstyle{plain}

\theoremstyle{plain}

\IEEEoverridecommandlockouts

\begin{document}

\title{Flexible-Position MIMO for Wireless Communications: Fundamentals, Challenges, and Future Directions}

\author{Jiakang~Zheng, Jiayi~Zhang, Hongyang~Du, Dusit~Niyato,~\IEEEmembership{Fellow,~IEEE}, Sumei~Sun,~\IEEEmembership{Fellow,~IEEE}, Bo~Ai,~\IEEEmembership{Fellow,~IEEE}, and Khaled B. Letaief,~\IEEEmembership{Fellow,~IEEE}
%
%
\thanks{J. Zheng, J. Zhang, and B. Ai are with the School of Electronic and Information Engineering and Frontiers Science Center for Smart High-speed Railway System, Beijing Jiaotong University, and J. Zheng is also with Nanyang Technological University; H. Du, and D. Niyato are with Nanyang Technological University; S. Sun is with Agency for Science, Technology and Research, Singapore; K. B. Letaief is with Hong Kong University of Science and Technology.}
}

\maketitle
\vspace{-1.8cm}
\begin{abstract}
The flexible-position multiple-input multiple-output (FLP-MIMO), such as fluid antennas and movable antennas, is a promising technology for future wireless communications. This is due to the fact that the positions of antennas at the transceiver and reflector can be dynamically optimized to achieve better channel conditions and, as such, can provide high spectral efficiency (SE) and energy efficiency (EE) gains with fewer antennas.
In this article, we introduce the fundamentals of FLP-MIMO systems, including hardware design, structure design, and potential applications. We shall demonstrate that FLP-MIMO, using fewer flexible antennas, can match the channel hardening achieved by a large number of fixed antennas. We will then analyze the SE-EE relationship for FLP-MIMO and fixed-position MIMO.
Furthermore, we will design the optimal trajectory of flexible antennas to maximize system sum SE or total EE at a fixed travel distance of each antenna.
Finally, several important research directions regarding FLP-MIMO communications are presented to facilitate further investigation.
\end{abstract}


\IEEEpeerreviewmaketitle

\section{Introduction}

To keep up with the exponential mobile traffic growth rate while simultaneously providing ubiquitous connectivity, industry and academic researchers are searching for new revolutionary wireless network technologies. The use of multiple antennas, also known as multiple-input multiple-output (MIMO) technology, is a primary solution to greatly improve the spectral efficiency (SE) and energy efficiency (EE) of wireless communications by adaptive beamforming and spatial multiplexing \cite{9113273}. In particular, the well-known massive MIMO system has become one of the key technologies of 5G due to its advantageous channel characteristics such as channel hardening and favorable propagation \cite{wong2022bruce}. For the upcoming 6G communication, MIMO still plays a crucial role and has derived a variety of novel technologies. For example, by deploying a larger number of antennas than massive MIMO in a compact space, the extremely large-scale MIMO can achieve significantly high degrees of freedom in near-field \cite{wang2023extremely}. A new distributed massive MIMO architecture called cell-free (CF) massive MIMO has also recently been developed to offer users seamless and uniform coverage with a large number of geographically distributed access points \cite{zheng2023mobile}. Obviously, these technologies all belong to fixed-position MIMO (FIP-MIMO), employing a large number of fixed antennas. Essentially, they improve performance at the cost of increasing hardware expenses. Especially for full-digital MIMO systems, each fixed antenna connected to a dedicated Radio Frequency (RF) transceiver is power hungry \cite{1367557}.

The concept of flexible-position MIMO (FLP-MIMO), which aims to match the gains of large-scale fixed antennas using fewer flexible ones, is currently being developed \cite{9770295,zhu2023movablemag}.
Compared with flexible MIMO altering many antenna characteristics \cite{wong2022bruce}, the basic principle of FLP-MIMO lies in the antennas at the transceiver and reflector can dynamically adjust their positions based on the surrounding environment.
Undeniably, compared with conventional FIP-MIMO that has been reliant on increasing the antenna number, the active adaptability of FLP-MIMO is a huge leap forward for multi-antenna systems. For this reason, FLP-MIMO has recently attracted significant academic attention, resulting in a variety of innovative approaches. An initial implementation of FLP-MIMO can be found in the switching antenna system, which turns on or off the antenna at different positions by controlling a radio frequency (RF) switch states based on channel conditions \cite{1367557}. It responds quickly, but still requires a considerable number of antennas and RF chains. Recently, a fluid antenna has been developed to enhance multiuser communication by relocating the cost-effective fluid radiator to positions experiencing minimum interference, thereby improving multiple access \cite{9650760}.
Then, mechanically driven movable antennas are explored in \cite{zhu2023movable}, which can be readily extended to 3D movement to fully exploit favorable propagation channels.
Moreover, antennas attached to mobile devices also qualifies as a form of FLP-MIMO systems \cite{zheng2023mobile}. {Compared to FIP-MIMO systems, the major advantages of the FLP-MIMO system are as follows:
\begin{itemize}
  \item FLP-MIMO can fully exploit the spatial degree of freedom by adjusting the antenna position, thereby achieving enhanced SE.
  \item FLP-MIMO optimizes channel gains using fewer flexible antennas, which reduces hardware requirements and improves EE.
  \item FLP-MIMO can be moved to positions where multi-user channels become asymptotically orthogonal, thus facilitating favorable propagation.
\end{itemize}}
Significantly, the beneficial feature of massive MIMO channel hardening can also be achieved with FLP-MIMO, as we will discuss and demonstrate in Section \ref{SE-EE}.

\begin{table*}[ht]
\centering
\caption{{Fundamentals of the FLP-MIMO system including its features, response speed, precision, hardware, cost, advantages, and disadvantages \cite{wong2022bruce,zheng2023mobile,1367557,9770295,zhu2023movablemag,9650760,zhu2023movable}.}}
\vspace{2mm}
\includegraphics[scale=0.6]{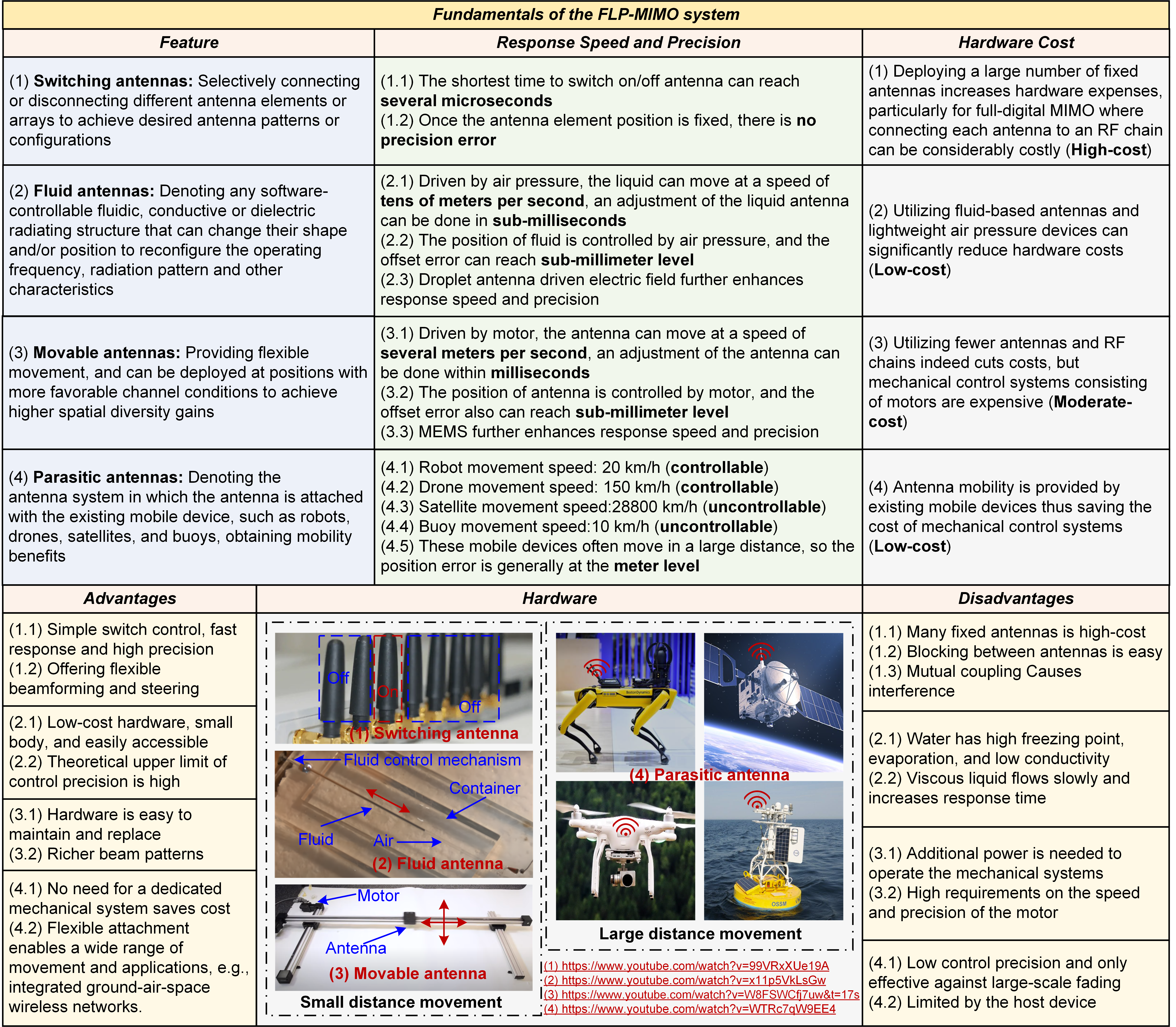}
\vspace{-3mm}
\label{fig:overview}
\end{table*}

Although FLP-MIMO systems offer several potential advantages, they are also having the following questions need to be answered:
\begin{itemize}
  \item Q1: What kind of hardware form and structure is used in the implementation of FLP-MIMO?
  \item Q2: How to evaluate the channel, rate and power consumption of FLP-MIMO?
  \item Q3: How to design optimal trajectories of antennas in FLP-MIMO?
\end{itemize}
This article will first provide an overview of the FLP-MIMO.
{The main contributions and responses to the aforementioned questions are summarized as follows:
\begin{itemize}
  \item We thoroughly examine the fundamentals of FLP-MIMO systems, highlighting its primary hardware implementations: switching antennas, fluid antennas, movable antennas, and parasitic antennas. We then explore their deployment structures and potential applications given practical requirements.
  \item We demonstrate that FLP-MIMO, despite utilizing fewer flexible antennas, can match the channel hardening accomplished by a large number of fixed antennas. Besides, we compare the SE-EE relationship between FLP-MIMO and FIP-MIMO.
  \item We formulate a trajectory optimization problem for FLP-MIMO aiming at maximizing system sum SE or total EE. Then, an AI-generated optimization method is adopted to find the optimal trajectory for the flexible antennas. Finally, we outline major future research directions.
\end{itemize}}

\section{Fundamentals of FLP-MIMO}

\begin{figure*}[ht]
\centering
\includegraphics[scale=0.62]{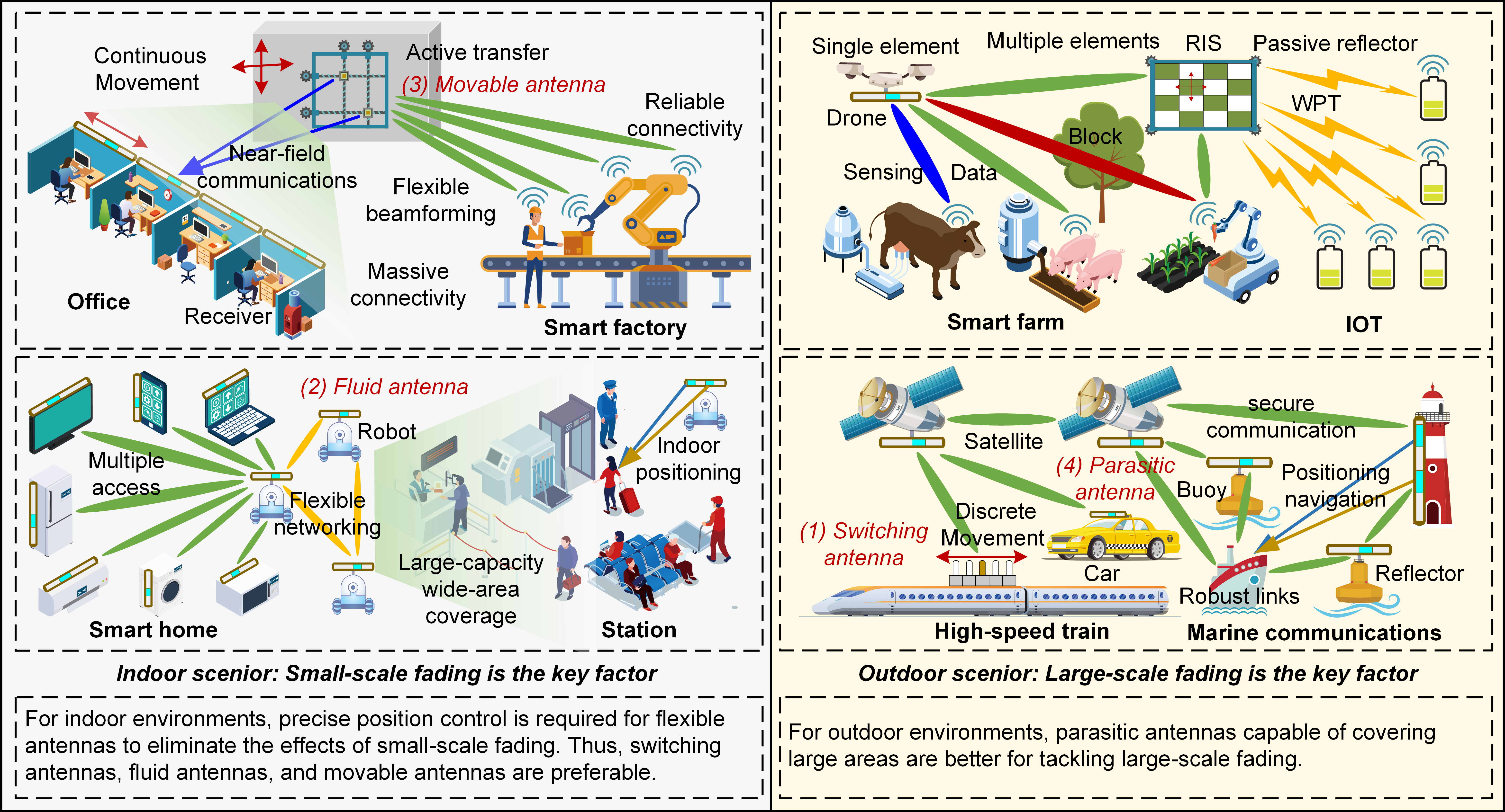}
\caption{{Potential applications of the FLP-MIMO systems \cite{9770295,zhu2023movablemag,zheng2023mobile}. In some cases, using switching antenna, fluid antenna, movable antenna, and parasitic antenna in a hybrid mode can be beneficial (e.g., attaching fluid antennas to drones).}}
\vspace{-3mm}
\label{fig:structure}
\end{figure*}

In this section, we introduce four hardware implementations of FLP-MIMO, including switching antennas, fluid antennas, movable antennas, and parasitic antennas, which are compared in TABLE~\ref{fig:overview}. Besides, we explore their structure designs and potential applications, as illustrated in Fig.~\ref{fig:structure}. Note that, these flexible antennas enable better coverage and adaptability to changing environmental conditions. By strategically relocating the antenna, it becomes possible to mitigate signal fading, blockage, and interference caused by various obstacles and scatterers. Therefore, the FLP-MIMO can reap the full diversity in the given spatial region.

\subsection{Hardware Design}

\subsubsection{Switching antennas}

The switching antennas use RF switches to select a suitable subset of antennas \cite{1367557}. Switching antennas operate based on the principle of selectively connecting or disconnecting different antenna elements or arrays to achieve desired antenna patterns or configurations.
This flexibility offers several advantages.
The switches can be activated or deactivated within microseconds, allowing for rapid reconfiguration of the antenna system.
Switching antennas also offer flexibility in beamforming and steering capabilities. By selectively connecting antenna elements, the system can form directional beams, adjust beamforming weights, and steer the antenna radiation pattern in desired directions.
However, switching antennas also have some disadvantages.
In switching antennas, the proximity of antenna elements can cause mutual coupling effects, leading to interference and degradation in the performance. Moreover, the switching antenna system faces scalability challenges due to the high cost and power consumption associated with connecting each antenna array to a dedicated radio frequency transceiver.

\subsubsection{Fluid antennas}

Fluid antenna refers to any software-controllable fluidic, conductive or dielectric radiating structure that can change their shape and/or position to reconfigure the operating frequency, radiation pattern and other characteristics \cite{9264694}.
Generally, fluid antennas are constructed using liquid metals or oil-based solutions, which is low-cost, eco-friendly, and easily accessible and integrated. Besides, fluid antennas can leverage high-resolution control mechanisms to achieve exceptional precision in tuning the position of the fluid radiator. This fine-grained control empowers the antenna to make precise adjustments, allowing for optimized performance and enhanced flexibility in meeting specific requirements.
However, fluid antennas face challenges related to the properties of the chosen fluid. For instance, water is not considered an ideal candidate for fluid antennas due to its high freezing point, tendency to evaporate, and relatively low conductivity.
Moreover, the viscosity of liquid metal and oil used in fluid antennas can affect the flow speed and response time \cite{9770295}. The longer response time may introduce delays in adjusting the shape or position of the antenna, potentially limiting their use in fast-changing communication environments.
{It is worth noting that a novel form of fluid antenna utilizes an electric field to control the position of the droplet antenna on a 2D/3D substrate, thereby significantly enhancing the beam pattern and reducing response speed.}

\subsubsection{Movable antennas}

Different from conventional fixed antennas that undergo random wireless channel variation, the movable antennas with the capability of flexible movement can be deployed at positions with more favorable channel conditions to achieve higher spatial diversity gains \cite{zhu2023movable}.
However, the implementation of movable antennas necessitates the integration of intricate mechanical systems, encompassing motors, gears, and control mechanisms, making the system bulky.
This complexity not only increases costs but also introduces potential points of failure. Besides, the movement and repositioning of movable antennas require additional power to operate the mechanical components. This can result in increased power consumption, which may be a concern in energy-constrained applications or battery-powered devices. Moreover, movable antennas require fast and precise operation of mechanical systems to meet the demands of real-time applications or scenarios.
Furthermore, the inclusion of mechanical components in movable antennas can increase their size, weight, and physical footprint. This can limit their applicability in compact devices or space-constrained environments where size and form factor are critical.
{Fortunately, most of the aforementioned challenges can be addressed through the utilization of micro-electromechanical systems (MEMS) to obtain high precision, miniaturization, and low power consumption, rendering movable antennas highly promising.}

\subsubsection{Parasitic antennas}

A parasitic antenna refers to the antenna system in which the antenna is attached with the existing mobile device to obtain mobility benefits.
{Hence, it is especially well-suited for integration into ground-air-space wireless networks.}
For instance, antennas can be integrated with easily controllable mobile devices like robots and drones, as well as with more challenging devices such as satellites and buoys to enhance significantly their communication range and operational efficiency \cite{zheng2023mobile}.
In contrast to the aforementioned three types of antenna systems, the parasitic antenna does not require a dedicated mechanical system for its design, consequently reducing hardware complexity and costs. Besides, it can achieve a broader range, shorting the distance between transceivers quickly, and significantly mitigating large-scale fading.
However, due to the limitations in controlling the precise positioning by the mobile device, the parasitic antenna often faces challenges in mitigating the impact of small-scale fading, such as deep fading.
Additionally, as each mobile device having its unique task and movement requirements, the research to determine the optimal position of parasitic antennas considering these constraints becomes a crucial direction.

\subsection{Structure Design}

\subsubsection{Signal Element and Multiple Elements}

Depending on particular needs and applications, FLP-MIMO systems can be composed of single or multiple elements. The benefit of a single element is simplicity, including position design, system installation, and movement control. However, it does not allow simultaneous propagation of multipath, so multiple elements are required to achieve better performance.
{With multiple flexible antennas, advanced signal processing techniques such as richer beams can also be implemented.}
However, the system will be more complex, as more complicated algorithms will be needed to optimize the position of multiple elements, and the cost of mechanical hardware is also higher. Therefore, there is a performance and cost trade-off in the selection of the number of elements \cite{bjornson2017massive}.

\subsubsection{Continuous Movement and Discrete Movement}

The difference between continuous and discrete movement mainly lies in the level of control accuracy of the mechanical system with FLP-MIMO. Continuous movement implies that, given sufficiently high-resolution control of the mechanism, the position of the antenna element in a flexible antenna can be theoretically fine-tuned with nearly limitless precision, mainly represented by fluid and movable antennas. Besides, discrete movement refers to those which can only transition among set positions or, due to accuracy constraints, can only be adjusted to predefined positions. This is primarily observed in switching and parasitic antennas.
In essence, a trade-off exists between the performance of the communication system and the accuracy of the mechanical system. While continuous movement provides superior spatial resolution, it demands more complex and accurate hardware. Conversely, discrete movement, being simpler and less mechanically demanding, might result in reduced communication performance.

\subsubsection{Active Element and Passive Element}

An active antenna element denotes a part of the antenna that has a direct connection to the transmitter or receiver, playing a crucial role in transmitting and/or receiving signals. On the other hand, a passive antenna element, while not being directly connected to the transmitter or receiver, contributes to modifying the antenna's radiation pattern.
{It serves like a passive reconfigurable intelligent surface (RIS) to reflect, refract, or diffract the signal, thereby altering the trajectory of the transmitted signal or the direction from which the signal is received \cite{9570143}. It is evident that the integration of RIS and FLP-MIMO benefits from each other, allowing for more flexible operations of the electromagnetic environment by changing position, phase and amplitude.} Therefore, based on the active element and the passive element, FLP-MIMO can act as a transceiver and reflector, respectively.

\vspace{-3mm}
\subsection{Potential Applications}

\subsubsection{Indoor Applications}

The electromagnetic environment within indoor spaces is generally more complex. Many scatterers, such as walls and furnishings, can cause significant multipath propagation, leading to deep channel fading, while at the same time acting as barriers that can block the direct path of the signal. Moreover, a large number of electronic devices typically found in modern indoor areas often produce significant RF pollution, making the process of signal transmission even more challenging.
In response to these challenges, larger antenna arrays are employed to produce highly directive beams, such as extremely large-scale MIMO operating in near-field communication \cite{wang2023extremely}. However, this also means greater hardware expenses and power consumption.
Utilizing fewer antennas, FLP-MIMO systems can provide flexible and adaptable solutions for indoor wireless communication by adjusting the position of the antenna.
For example, deploy the FLP-MIMO on one side of the iPad, and by moving the receiving antenna to a position that minimizes deep fading and RF interference, the iPad can always receive the best wireless communication service. In addition, FLP-MIMO can be deployed on a wall, enabling a small range of antenna movement to enhance the probability of establishing a direct link. Therefore, the FLP-MIMO can be deployed in various indoor scenarios, such as shopping malls, stadiums, and convention centers, to improve the quality of the signal.

\subsubsection{Outdoor Applications}

Outdoor communication systems are commonly designed to provide reliable and efficient wireless coverage over large areas. To realize this aim, a greater number of access points are typically employed outdoors, such as cell-free massive MIMO systems. They offer seamless coverage to users through joint beam management, combating the impacts of path loss, shadow fading, and interfering signals.
Unlike cell-free massive MIMO, which utilizes as many fixed access points as possible to maximize the degree of freedom, FLP-MIMO systems can employ a smaller number of movable access points to capitalize on the degrees of freedom across the entire region. Hence, FLP-MIMO has a very wide range of applications outdoors. For instance, the mobile robot carries antennas to shorten the distance between the transceivers to reduce the path loss. In addition, drones carry antennas to avoid obstacles and provide effective and reliable line-of-sight wireless communication services. Besides, in multi-user scenarios, flexible antennas can jointly alter their positions to achieve the optimal position with the least interference. Moreover, employing FLP-MIMO to create the most favorable channel propagation conditions and thereby maximize the efficiency of wireless power transfer represents another significant application.

FLP-MIMO achieves notable performance improvements by relocating the antenna to positions less impacted by both small-scale and large-scale fading. However, this might increase hardware cost and power consumption as well as complicate modelling and algorithms, so careful balancing is essential according to different scenarios.

\section{Channel Hardening and Spectral-Energy Efficiency}\label{SE-EE}

It is well known that channel hardening is an important property of massive MIMO channels benefiting from spatial diversity. Specifically, channel hardening means that the beamforming transforms the fading multi-antenna channel into an almost deterministic scalar channel. This property alleviates the need for avoiding small-scale fading and simplifies the required signal processing and resource allocation.
For example, channel hardening can often eliminate the need for channel estimation at the terminals, and the associated transmission of pilots \cite{bjornson2017massive}.
Moreover, there is no need to adapt the power allocation or scheduling to small-scale fading variations, and linear processing is nearly optimal \cite{bjornson2017massive}. However, maintaining this characteristic necessitates a substantial number of antennas, which incurs significant costs.
{Fortunately in actual deployments, FLP-MIMO can achieve the same degree of channel hardening as a large number of fixed antennas with a small number of flexible antennas by moving the antenna elements to the positions with optimal channel conditions, which was pointed out in \cite{wong2022bruce}.}
{Next, we verify this benefit and analyze the key factors as follows.}

\begin{figure}[ht]
\centering
\includegraphics[scale=0.48]{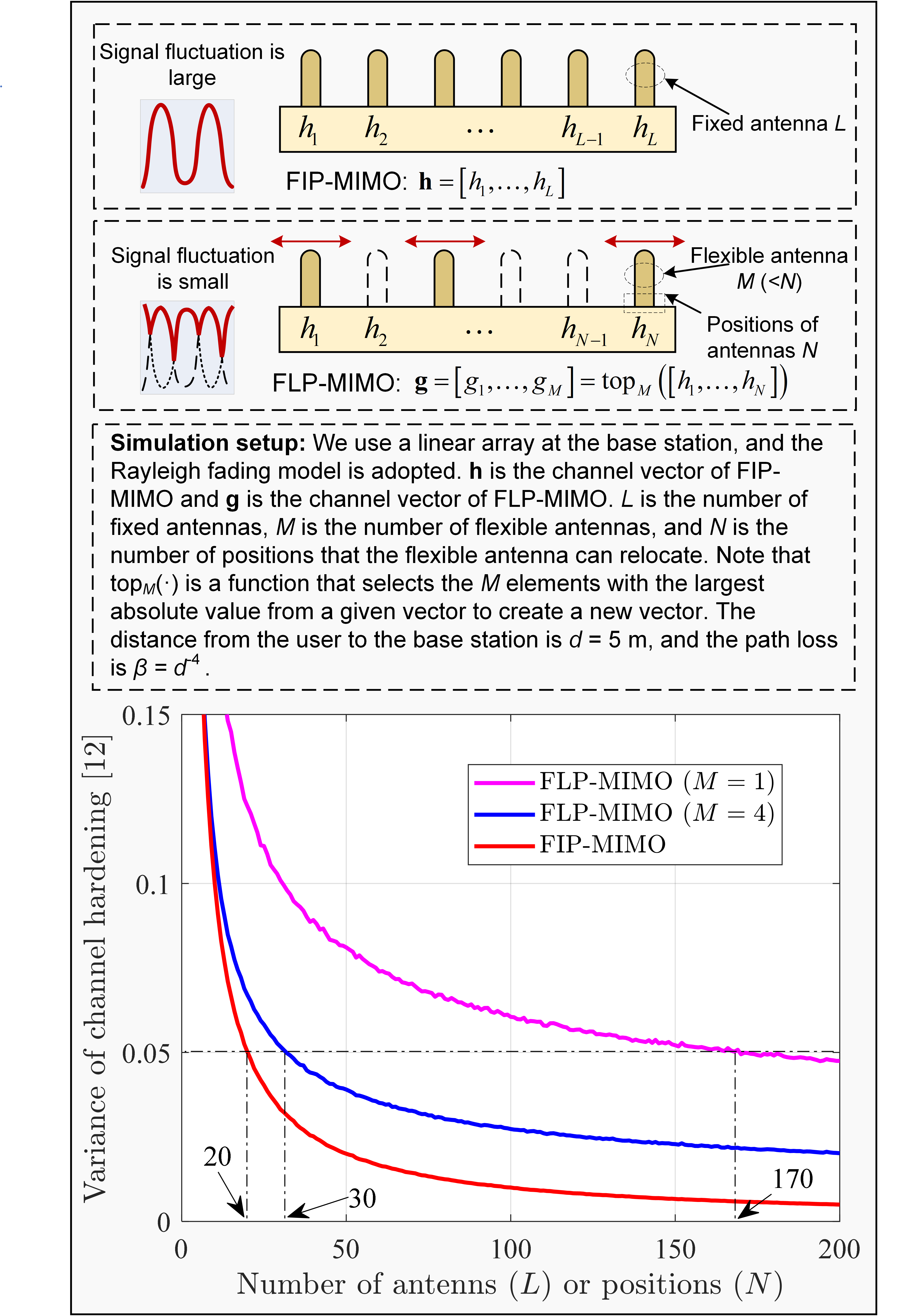}
\caption{{Variance of channel hardening against the number of antennas of FIP-MIMO and the number of positions of FLP-MIMO. The variance of channel hardening is defined as \cite[Eq. (2.17)]{bjornson2017massive}, and the smaller value of it, the stronger channel hardening.}}
\vspace{-3mm}
\label{fig:ChannelHardening}
\end{figure}

Considering a linear array such as switching antennas and fluid antennas, Fig.~\ref{fig:ChannelHardening} shows the variance of channel hardening against the number of antennas in FIP-MIMO and the number of positions in FLP-MIMO. It is clear that the variance of channel hardening of both FLP-MIMO and FIP-MIMO increases with the increase of the number of positions and the number of antennas, respectively. However, their principles are different, where FIP-MIMO uses the \textbf{law of large numbers} to achieve strong channel hardening, and FLP-MIMO uses \textbf{extreme value theory} to achieve strong channel hardening. In the simulation, we can also find that FLP-MIMO with a single antenna can utilize $170$ positions to achieve the same level of channel hardening as FIP-MIMO with $20$ antennas. Furthermore, FLP-MIMO with $4$ independent antennas requires only $30$ positions to achieve the same level of channel hardening, which is a huge hardware savings compared to $20$ fixed antennas.
{It can be concluded that when the degree of channel hardening of FLP-MIMO and FIP-MIMO is the same, the number of fixed antennas falls between the number of flexible antennas and the number of positions that the flexible antenna can relocate.}
It is worth noting that it is necessary for the antenna in the FLP-MIMO system to move correctly and quickly to the position with the optimal channel condition. Thus, how to reach the optimal position is an important research direction, which will be discussed in Section~\ref{optimal location}.

\begin{figure}[h]
\centering
\includegraphics[scale=0.7]{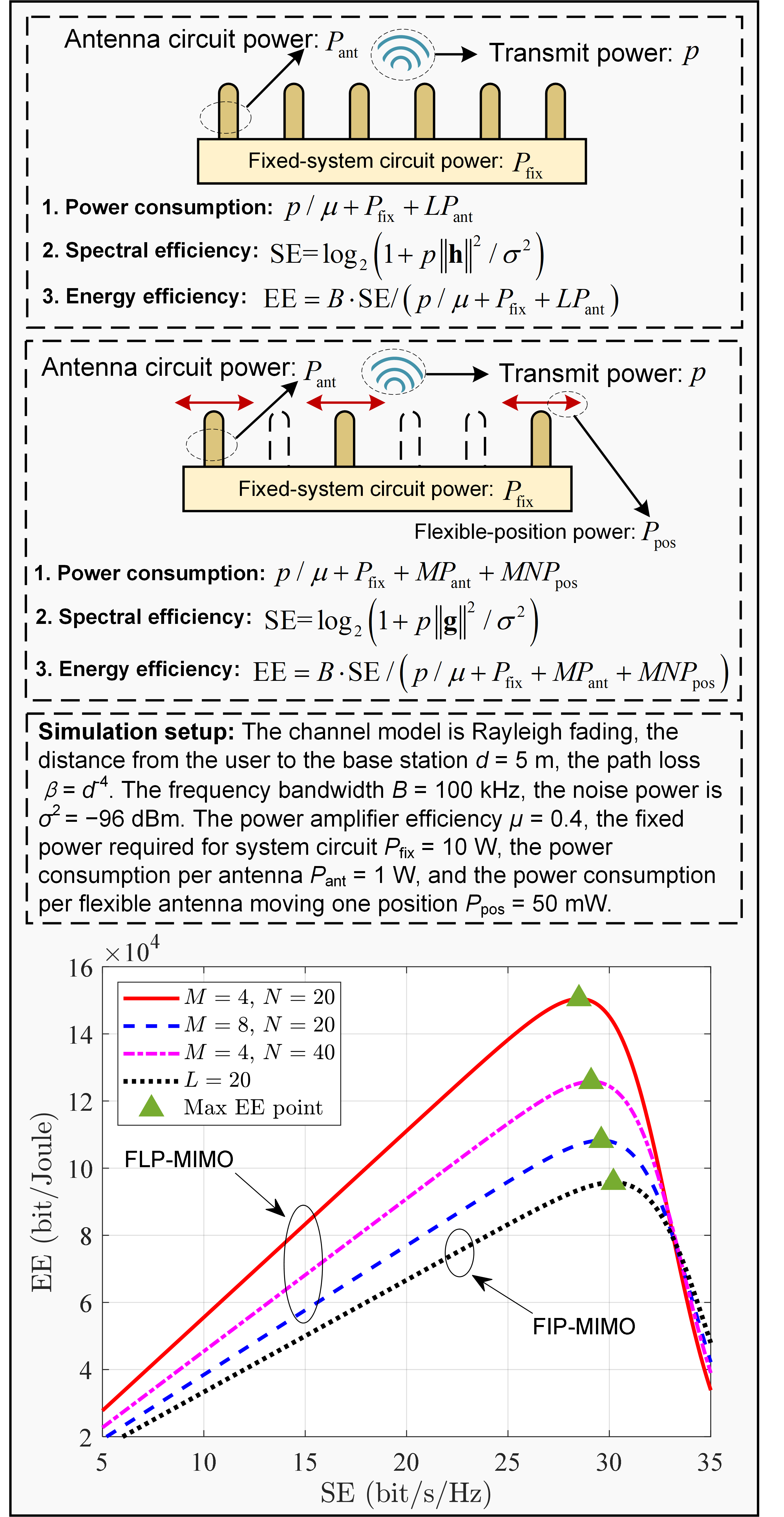}
\caption{{SE and EE relationship for FLP-MIMO and FIP-MIMO.}}
\vspace{-3mm}
\label{fig:EE_SE}
\end{figure}

Next, we investigate the SE and EE of FLP-MIMO systems. SE is defined as the number of bits that can be successfully transmitted per unit of time and bandwidth (bit/s/Hz), and EE is defined as the number of bits that can be successfully transmitted per unit of energy (bit/Joule) \cite{bjornson2017massive}.
{Clearly, the predictable and stable FLP-MIMO channel due to strong channel hardening can achieve higher SE as the deterministic channel enables the system to support high data rates over given bandwidth, thus efficiently utilizing the available spectrum as shown in Fig.~\ref{fig:EE_SE}.}
For deriving the total EE of the FLP-MIMO system, we further need its accurate power consumption model, which can be divided into several components:
\begin{itemize}
\item \textit{Effective Transmit Power:} This is the effective power required to transmit signals from the antennas to the receivers. Note that the transmit power does not represent the effective transmit power needed for transmission since the former does not account for the efficiency of the power amplifier. The effective transmit power here includes radiated transmit power and power amplifier dissipation. Therefore, the effective transmit power depends on the power level of the signals, the number of antennas, and the efficiency of the power amplifier. Bedsides, the efficiency of a power amplifier is defined as the ratio of output power to input power. When the efficiency is low, a large portion of the input power is dissipated as heat.
\item \textit{Circuit Power:} This is the power required for the circuit consumption of the whole system, detailed including the power consumed by the transceiver chains, the channel estimation process, the channel encoding and decoding units, the load-dependent backhaul signaling, the signal processing at the base station, and a fixed power required for control signaling and baseband processors \cite{bjornson2017massive}. Simply, it can be modeled as a fixed system circuit power, and the transceiver circuit power, with the latter being dependent on the number of antennas.
    Note that the accurate circuit power consumption model for transceiver hardware at the base station and at the user is necessary for accurate and practical system optimization.
\item \textit{Flexible-position Power:} This is a unique component for FLP-MIMO systems. It is the power required to move the antennas to different positions. The amount of this power depends on factors such as the number of antennas that need to move, the distance of the movement, the speed of the movement, and the physical characteristics of the antennas (e.g., weight and shape).
    Based on the velocity-dependent propulsion energy model for drones \cite{7888557}, we provide a preliminary analyzable flexible-position power model.
    To ensure that the time for the flexible antenna to complete the movement is constant, then the power required by each flexible antenna increases linearly with the number of positions.
    Thus, we consider that the power cost per antenna unit movement distance modeled as a fixed value, and the flexible-position power is a linear function of the number of antennas and the movement distance.

\end{itemize}
By understanding these different components of power consumption, one can optimize the FLP-MIMO system to improve its energy efficiency. For example, by enhancing the antenna movement mechanism, the power needed for antenna position adjustment can be reduced. Similarly, the power used for signal processing can be reduced through the optimization of signal processing algorithms.

\begin{figure*}[ht]
\centering
\includegraphics[scale=0.82]{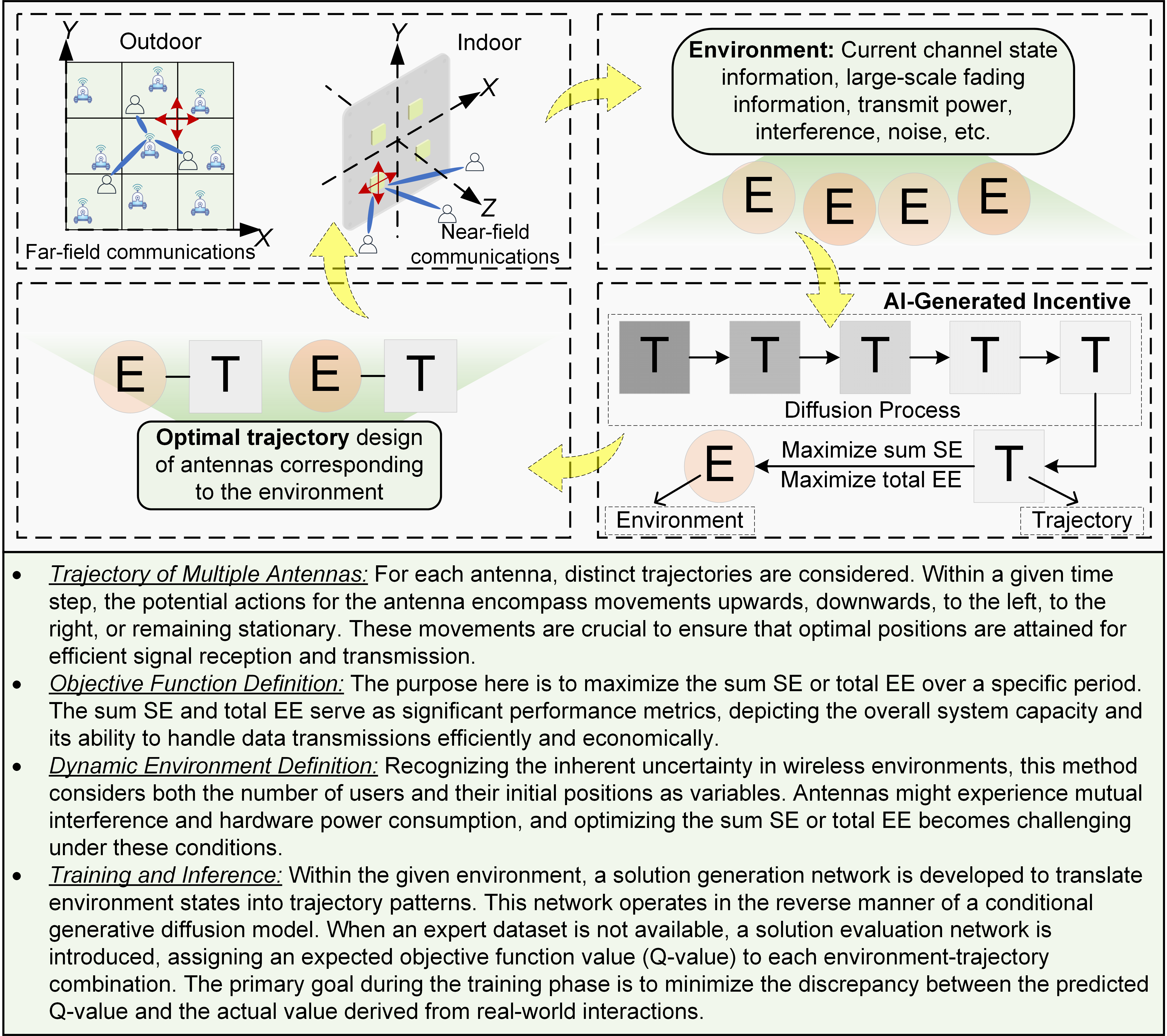}
\caption{AI-generated optimization technology for the trajectory design of FLP-MIMO systems. We aim to find trajectories that optimize the sum SE or total EE of systems with each antenna traveling a fixed distance. Through real-time parameter input and result output of the AI-generated module, the optimal trajectory can be adaptively fine-tuned according to the dynamic environment.}
\vspace{-3mm}
\label{fig:trajectory_in_out_door}
\end{figure*}

In Fig.~\ref{fig:EE_SE}, we perform a performance comparison between FLP-MIMO and FIP-MIMO by showing the dynamic relationship between SE and EE.
{It is clear that the FLP-MIMO system achieves the larger EE than that of the FIP-MIMO system. The reason is that each antenna connected to a dedicated RF transceiver, is very cost-prohibitive and power-hungry.}
In contrast, moving a small antenna with a distance of one position consumes less power.
{By observing the maximum EE point, we find that increasing the number of antennas and adjusting the positions in the FLP-MIMO system increase SE but reduce the maximum EE as more power is consumed, and thus their trade-off requires careful decision in actual deployment.}
Importantly, FLP-MIMO with fewer flexible antennas can achieve larger SE than that of the FIP-MIMO with large number of antennas by increasing the number of positions, but at the expense of more power consumed by antenna movement.

Apart from mitigating small-scale fading, another significant and important advantage of the FLP-MIMO system is its capability for flexible coverage. For instance, it can adaptively modify its position relative to the user, taking into account factors such as path loss, obstacles, and interference conditions \cite{zhu2023movable}. Therefore, the practical enhancements in SE and EE of the FLP-MIMO system could exceed our simulation results only considering small-scale fading. Thus, these simulations should be viewed as a lower bound.

\begin{figure*}[ht]
\centering
\includegraphics[scale=0.52]{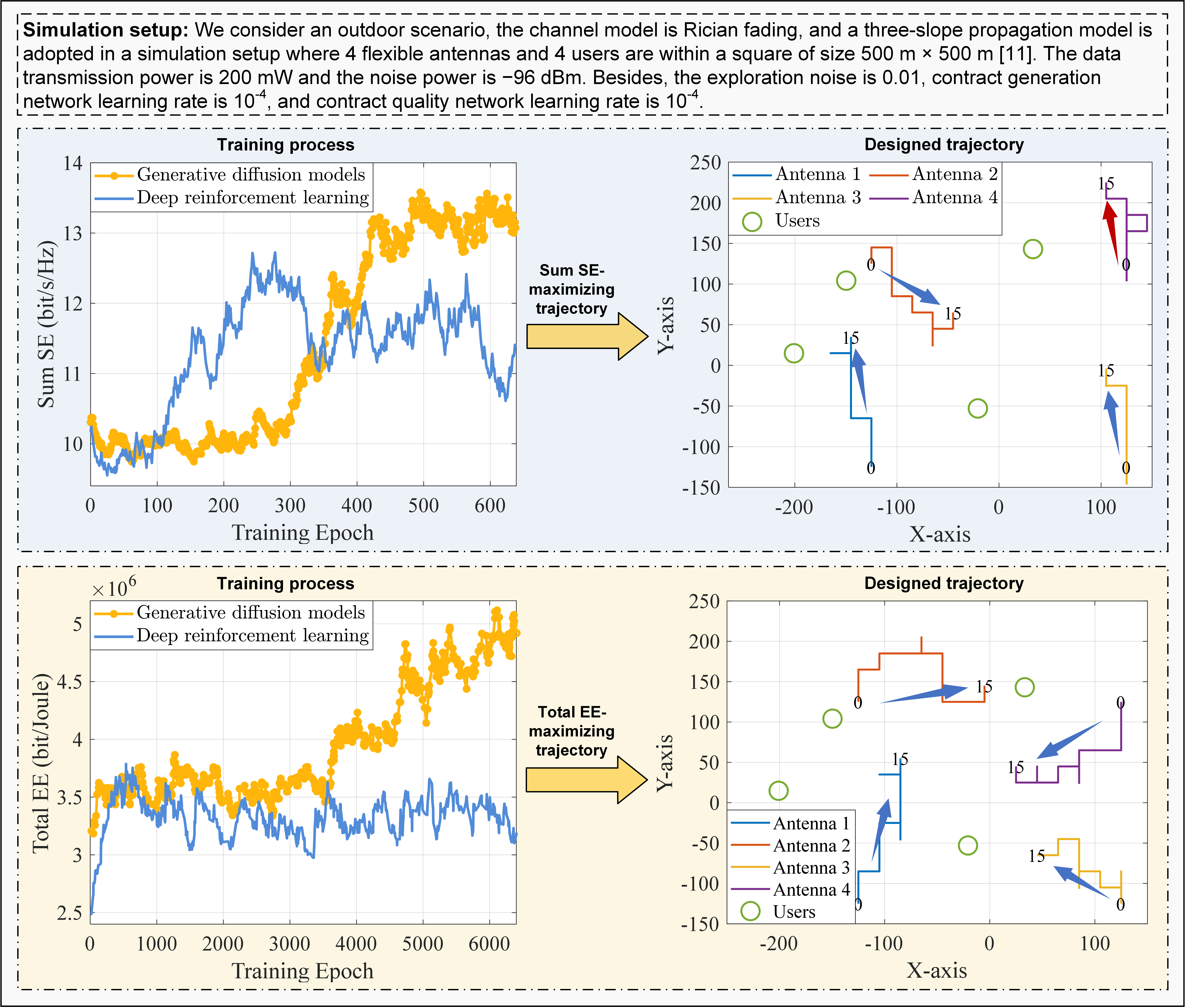}
\caption{{Trajectory design of FLP-MIMO with an AI-generated optimization method. The experimental platform is built on a Windows 11 system with 12th Gen Intel(R) Core(TM) i7-12700H CPU and NVIDIA GeForce RTX 3060 GPU.}}
\vspace{-3mm}
\label{fig:trajectory}
\end{figure*}
\section{Trajectory Design with AI-Generated Optimization}\label{optimal location}

As per the analysis in Section~\ref{SE-EE}, it is evident that the substantial SE gain of the FLP-MIMO system relies heavily on accurately relocating the antenna to a position with optimal channel conditions. Besides, efficient trajectory design is crucial for reducing power consumption and enhancing SE and EE.
Therefore, we explore trajectory optimization for multiple flexible antennas in a FLP-MIMO system with multiple users as shown in Fig.~\ref{fig:trajectory_in_out_door}.
{Unlike the study in \cite{zhu2023movable} which optimizes the position of the movable antenna to minimize the transmit power of the user, we focus on trajectories that maximize system sum SE or total EE at a fixed travel distance of each antenna.}
There are many ways to solve this problem, such as convex optimization theory \cite{zhu2023movable}, deep reinforcement learning \cite{10158526}. However, the high state dimension makes the former complex and makes the latter easy to fall into local optimum.

Therefore, we turn to a generative AI technique, called AI-generated optimization, whose key technology is generative diffusion models \cite{du2023beyond}.
From an initial input, the generative diffusion model gradually adds Gaussian noise over multiple steps, termed the forward diffusion process, generating targets for the denoising neural network. Following this, the neural network undergoes training to reverse the noising process and recover the data. The reverse diffusion method enables the production of new data. Beneficially, generative diffusion models have a strong generative capacity, ideal for dynamic network optimization. Besides, generative diffusion models can readily integrate conditioning data into the denoising process.
Next, we propose a trajectory design method leveraging AI-generated optimization.
The primary objective of this trajectory design is to maximize the sum SE or total EE of systems in dynamic wireless environments.
Specifically, this method consists of four main steps: \emph{Trajectory of Multiple Antennas}, \emph{Objective Function Definition}, \emph{Dynamic Environment Definition}, and \emph{Training and Inference}, as illustrated in Fig.~\ref{fig:trajectory_in_out_door}.

As shown in Fig.~\ref{fig:trajectory}, we compare the trajectory design of FLP-MIMO with AI-generated optimization and deep reinforcement learning methods. The training process reveals that optimization with deep reinforcement learning has a rapid performance improvement but then falls into local optimums, especially for the case of total EE with more complex solution space. Interestingly, both sum SE and total EE based on AI-generated optimization can converge to the optimum continuously and smoothly. The reason is that diffusion models achieve better sampling quality and has better long-term dependence processing capability \cite{10158526,du2023beyond}.
Besides, from the designed trajectory for maximum sum SE, we find that antenna 1, antenna 2 and antenna 3 are moving to better cover all users, while antenna 4 is moving towards the edge. The reason is that multi-user interference is more severe on antenna 4, so it is marginalized to maximize sum SE performance. For the designed trajectory maximizing total EE, all antennas move towards the center of the users for a trade-off between SE and EE. This also confirms the result in Fig.~\ref{fig:EE_SE} that the SE at the peak EE is not optimal.

\section{Future Directions}

\subsection{Resource Allocation}

Resource allocation optimization is a key approach for further improving the performance of FLP-MIMO systems. Due to the change of the network state caused by the movement of the antenna in FLP-MIMO, the traditional resource allocation method is no longer applicable.
{Therefore, dynamic beamforming, combining, power control and other resource allocation algorithms need to be jointly designed with the position of the flexible antenna.}
These resource allocation techniques requires robust and dynamic algorithms that can adapt to the constantly changing position of the antennas.
{Particularly when the FLP-MIMO operates indoors, it is crucial to exploit resource allocation under near-field communications with Doppler effect.}

\subsection{Integrated Sensing and Communications}

With the overlapping of communication and radar sensing frequency bands, integrated sensing and communications have become the key technologies for mutual benefits with limited resources.
Thanks to integrated sensing and communications that give real-time information about the environment, the FLP-MIMO can move around to avoid interference and obstacles, making sure that wireless transmission is fast and reliable. In turn, FLP-MIMO increases the probability of line-of-sight transmission and enhances the sensing capacity. Moreover, completing complex localization and sensing tasks using only a single flexible antenna is also an interesting research direction.

\subsection{Semantic Communications}

Semantic communication is a promising technology for future wireless data transmission because it sends only task-relevant semantic information. FLP-MIMO can fully capture the spatial characteristics, but this also means it has to handle a vast amount of complex, high-dimensional data.
Instead of sending the original source data, semantic communication-assisted FLP-MIMO compresses or encodes it, lowering the quantity of data that must be transmitted while also reducing the required bandwidth and power consumption.
Furthermore, a fascinating area of study is the trajectory design in FLP-MIMO systems based on the value of semantic information.

\section{Conclusions}

In this article, our focus is on the fundamentals of FLP-MIMO, a crucial technology for the next-generation wireless communication systems. We explored their hardware implementations including switching antennas, fluid antennas, movable antennas, and parasitic antennas, as well as their structure design and potential applications. Next, the potential of FLP-MIMO for achieving channel hardening, even with a limited number of flexible antennas, is discussed and verified. Besides, the power consumption model of FLP-MIMO is provided to compare its SE and EE with FIP-MIMO.
An AI-generated optimization method is also employed to determine the optimal trajectory for the movement of FLP-MIMO.
Finally, we presented promising future research directions and hope this article will be helpful in designing and implementing FLP-MIMO for future wireless communications.

\vspace{0cm}
\bibliographystyle{IEEEtran}
\bibliography{IEEEabrv,Ref}

\section*{Biographies}

\textbf{Jiakang Zheng} is currently pursuing the Ph.D. degree at the School of Electronic and Information Engineering, Beijing Jiaotong University, Beijing, China. His research interests include cell-free massive MIMO and performance analysis of wireless systems (e-mail: jiakangzheng@bjtu.edu.cn).

\textbf{Jiayi Zhang} [SM'20] is a professor with the School of Electronic and Information Engineering, Beijing Jiaotong University, Beijing, China. His research interests include cell-free massive MIMO and reconfigurable intelligent surfaces (e-mail: jiayizhang@bjtu.edu.cn).

\textbf{Hongyang Du} is currently pursuing the Ph.D degree at the School of Computer Science and Engineering, Nanyang Technological University, Singapore. His research interests include semantic communications, resource allocation, and communication theory (e-mail: hongyang001@e.ntu.edu.sg).

\textbf{Dusit Niyato} [F'17] is a professor with the School of Computer Science and Engineering, Nanyang Technological University, Singapore. His research interests are in the areas of the Internet of Things (IoT), machine learning, and incentive mechanism design (e-mail: dniyato@ntu.edu.sg).

\textbf{Sumei Sun} [F'16] is a principal scientist, Deputy Executive Director (Research), and Head of the Communications and Networks Department at the Institute for Infocomm Research (I2R), Agency for Science, Technology, and Research (A*STAR), Singapore. She also holds a joint appointment with Singapore Institute of Technology and an adjunct appointment with the National University of Singapore, both as a full professor (e-mail: sunsm@i2r.a-star.edu.sg).

\textbf{Bo Ai} [F'22] is a professor with the State Key Laboratory of Rail Traffic Control and Safety, Beijing Jiaotong University, Beijing, China. His research interests include rail traffic mobile communications and channel modeling (e-mail: boai@bjtu.edu.cn).

\textbf{Khaled B. Letaief} [F'03] has been with HKUST since 1993 where he was the Dean of Engineering, and is now a Chair Professor and the New Bright Professor of Engineering. From 2015 to 2018, he was with HBKU in Qatar as Provost. He is an ISI Highly Cited Researcher. He has served in many IEEE leadership positions including ComSoc President, Vice-President for Technical Activities, and Vice-President for Conferences (e-mail: eekhaled@ust.hk).

\end{document}